\documentclass{vgtc}                          

\ifpdf
  \pdfoutput=1\relax                   
  \usepackage{graphicx}                
  \DeclareGraphicsExtensions{.pdf,.png,.jpg,.jpeg} 
\else
  \usepackage{graphicx}                
  \DeclareGraphicsExtensions{.eps}     
\fi%

\graphicspath{{figures/}{pictures/}{images/}{./}} 

\usepackage{microtype}                 
\PassOptionsToPackage{warn}{textcomp}  
\usepackage{textcomp}                  
\usepackage{mathptmx}                  
\usepackage{times}                     
\usepackage{cite}                      
\usepackage{tabu}                      
\usepackage{booktabs}                  

\onlineid{0}

\vgtccategory{Research}

\vgtcinsertpkg

\title{Boundaries, Extensions, and Challenges of Visualization for Humanities Data: Reflections on Three Cases}

\author{Rongqian Ma\thanks{e-mail: rm56@iu.edu}}
\affiliation{\scriptsize Indiana University Bloomington}

\abstract{This paper discusses problems of visualizing humanities data of various forms, such as video data, archival data, and numeric-oriented social science data, with three distinct case studies. By describing the visualization practices and the issues that emerged from the process, this paper uses the three cases to each identify a pertinent question for reflection. More specifically, I reflect on the difficulty, thoughts, and considerations of choosing the most effective and sufficient forms of visualization to enhance the expression of specific cultural and humanities data in the projects. Discussions in this paper concern some questions, such as, how do the multi-modality of humanities and cultural data challenge the understanding, roles, and functions of visualizations, and more broadly, visual representations in humanities research? What do we lose of the original data by visualizing them in those projects? How to balance the benefits and disadvantages of visual technologies to display complex, unique, and often culturally saturated humanities datasets?%
} 

\CCScatlist{ 
  \CCScat{Human-centered computing}{Visualization}{Visualization application domains}{Information visualization}
}

\begin{document}


\firstsection{Introduction}

\maketitle

Visual technologies have been widely used for digital humanities research. Visualization in the digital humanities sits at the intersection of the DH scholarship and that of the visualization, having attracted discussions from scholars in both fields. According to Bailey and Pregill (2014), ``twenty-first-century humanities scholars find themselves in the midst of a visualization renaissance of sorts with information analysis and visualization literacy recognized as fundamental skills in the academy" \cite{bailey_speak_2014}. The use of visualizations for interpretation and analysis has been advocated since the 2000s, especially in traditional humanities research areas such as literary studies where Moretti (2005) published his influential work titled \textit{Graphs, Maps, Trees: Abstract Models for a Literary History} \cite{moretti_graphs_2005}. A massive number of projects and visualization tools have been developed to assist with humanities inquiries and research exploration, covering various types of humanities data and analyses including the spatial visualization, temporal visualization, textual visualization,  and 3D visualizations \cite{munster_visual_2020}. From the literature, two interesting themes emerged: (1) the nature of data in the humanities research context; and (2) principles of humanistic visualizations. \par

\subsection{Nature of Humanities Data}

Discussion of \textit{humanities visualization} should first benefit from an in-depth exploration of the connotations of data for humanities subjects and topics. As illustrated by Schöch (2013), ``most of the colleagues in literary and cultural studies would not necessarily speak of their objects of study as data. If you ask them what it is they are studying, they would rather speak of books, paintings and movies; of drama and crime fiction, of still lives and action painting; of German expressionist movies and romantic comedy. … Maybe they would talk about what they are studying as texts, images, and sounds. But rarely would they consider their objects of study to be data'' \cite{schoch_big_2013}. However, the mass digitization of cultural and humanities materials has introduced new occasions, as therefore unique challenges, for scholars working with cultural materials in humanities-oriented research. From another perspective, this quote also aptly captures the unique characteristics of data in the humanities research context, which has been explored and discussed in existing scholarship.\par

For instance, Borgman (2010) discussed the unique characteristics of humanities data compared with those in scientific research contexts \cite{borgman_digital_2010}. Unlike natural or social scientists whose data usually come from experimental observations and are clearly different from ``publications,” humanities data are ``innumerable” and their boundaries with publications are ``fuzzy.” On one hand, ``publications and other documents are essential sources of data to humanists: Newspapers, unpublished correspondence, diaries, manuscripts, and photographs are among the most heavily used sources by academic historians, for example. They are analyzed for facts, evidence, themes, and interpretations.” On the other, ``almost any document, physical artifact, or record of human activity can be used to study culture. Humanities scholars value new approaches and recognizing something as a source of data (e.g., high school yearbooks, cookbooks, or wear patterns in the floors of public places) can be an act of scholarship.” \par

Humanities data are also distinctive from scientific data due to their ``dispersion and separation from context:" ``Cultural artifacts are bought and sold, looted in wars, and relocated to museums and private collections. International agreements on the repatriation of cultural objects now prevent many items from being exported, but items that were exported decades or centuries ago are unlikely to be returned to their original sites. Those who hold cultural artifacts create the records that describe them, and thus the records are also dispersed” \cite{borgman_digital_2010,geser_resource_2004}.\par

In addition to the unique forms and characteristics, humanities data are also distinct in terms of their value- and interpretation- laden nature. Drucker (2011) highlighted this aspect with the notion of \textit{capta} \cite{drucker_humanities_2011}. Compared with the widely known concept of data, which refers to ``things given” in Latin, \textit{capta} is a \textit{taken}, which captures the interpretive nature of humanities inquiries. This understanding of humanities data supports Drucker’s theories of a ``humanistic visualization,” which has also been shared and developed in the field.\par

\subsection{Principles of Humanistic Visualization}

Following the unique characteristics of humanities data, the next question is: How to design the best visualizations for humanities data? One important theme of discussion in the emerging scholarship on DH visualization has focused on what is the value of visualization in digital humanities research, and what a ``humanistic visualization” should look like and hypothetically be different from visualization in general (e.g., scientific visualization or information visualization). Jessop (2008) argued that visualization should be a scholarly activity in the digital humanities, not simply a technique or a tool, which means that they are highly interwoven into humanities inquiries and interpretation \cite{jessop_digital_2008}. Such embedded visualizations constitute the concept of ``humanistic visualization.” Manovich (2011) raised the concept of ``direct visualization,” a method that ``creates new visual representations from the actual visual media objects or their parts,'' without any reduction \cite{manovich_what_2011}. This view of visualization values complexity and the preservation of original forms of the humanities data, which may better serve the purposes of humanistic inquiries. \par

\subsection{Roadmap}

But how are these principles implemented in actual research cases? How do the ideals of humanities data representations fit in the current knowledge infrastructure? By discussing three case studies where visual representations of various kinds are implemented, this paper attempts to raise and tackle certain questions related to humanistic visualization: For example, how do we define the boundaries and extensions of a \textit{visualization} for humanities and cultural data? How to design the most powerful, effective, yet also functional visual representations for digital humanities projects? What difficulty and challenges do project curators face in the process? More specifically, I reflect on these issues with three case studies where visualization techniques have been utilized to enhance representation of the humanities data and corpus in the digital environment. Critical discussions of these case studies serve as a building block for future theoretical generalization and research of visualizations applied for humanities datasets.\par

\section{CASE STUDIES}
I draw on three projects in different stages of development that focus on three different forms of humanities data. In the first case study, I discuss an ongoing project that explores the visualization of the variety of village gazetteer data in the context of contemporary China. Following this study, the second project offers another perspective to think about roles and function of ``visualization,'' or more broadly, visual representations, with oral history data about China’s Cultural Revolution (1966-1976). The third and final study further pushes forward my thinking about visuals’ roles, especially questions such as if digitization could function as a form of “visualization” for archival data, and how well they help enhance the representation of archival materials, especially in the digital environment. 

\subsection{Visualizing Contemporary Village Data}

Village gazetteers have been an important genre of writing and valuable historical resource about the rural life and society in China. Usually compiled and published by local village committees and administrations, rather than official governments, village gazetteers contain raw, uncensored, and detailed data about various aspects of a specific village, including its natural climate and resources, population, economics, healthcare, and culture. Although there are many published village gazetteers, the digital presence of them has been quite rare, let alone the existence of a centralized digital collection of village gazetteer data. The Contemporary Chinese Village Gazetteer Data (CCVGD) project aims to bridge this gap by building an interactive search platform of contemporary Chinese village gazetteer data based on over 3000 published resources \cite{zhang_contemporary_2020}. This data platform includes a wide range of data categories, such as villages’ administrative information, natural environment, population, economics, healthcare, and education. Majority are numeric data, offering means and invaluable information for social science research of contemporary China. \par

\begin{figure}[tb]
 \centering 
 \includegraphics[width=\columnwidth]{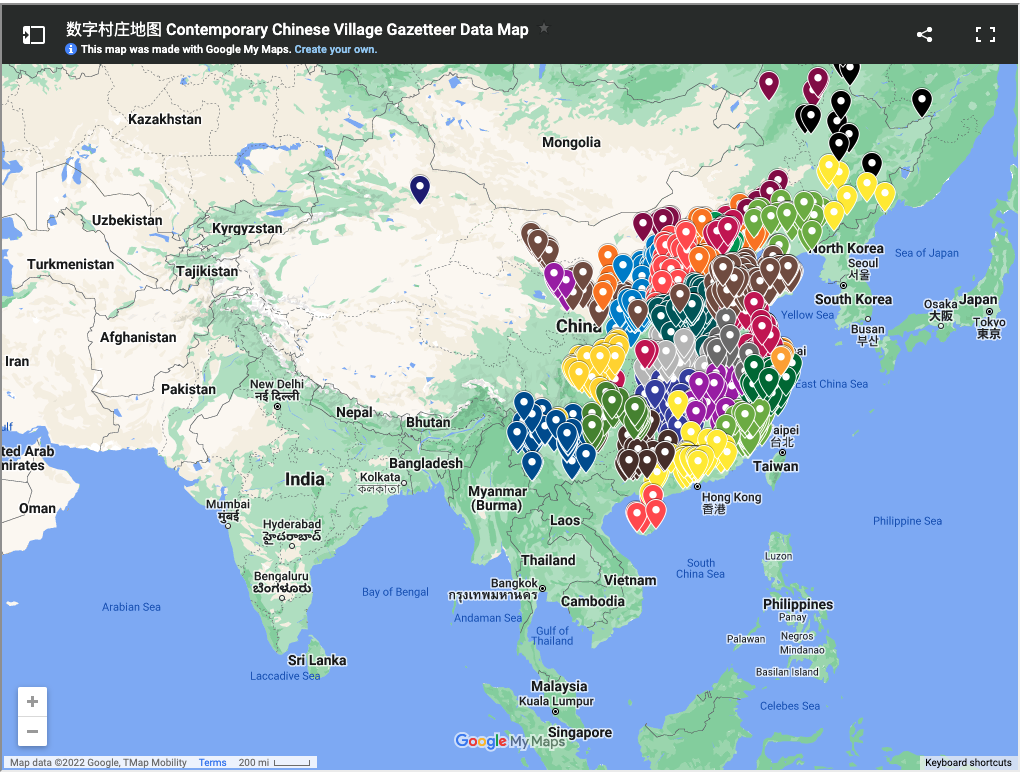}
 \caption{Map for the CCVG data project} 
 \label{fig:1}
\end{figure}

\begin{figure}[tb]
 \centering 
 \includegraphics[width=\columnwidth]{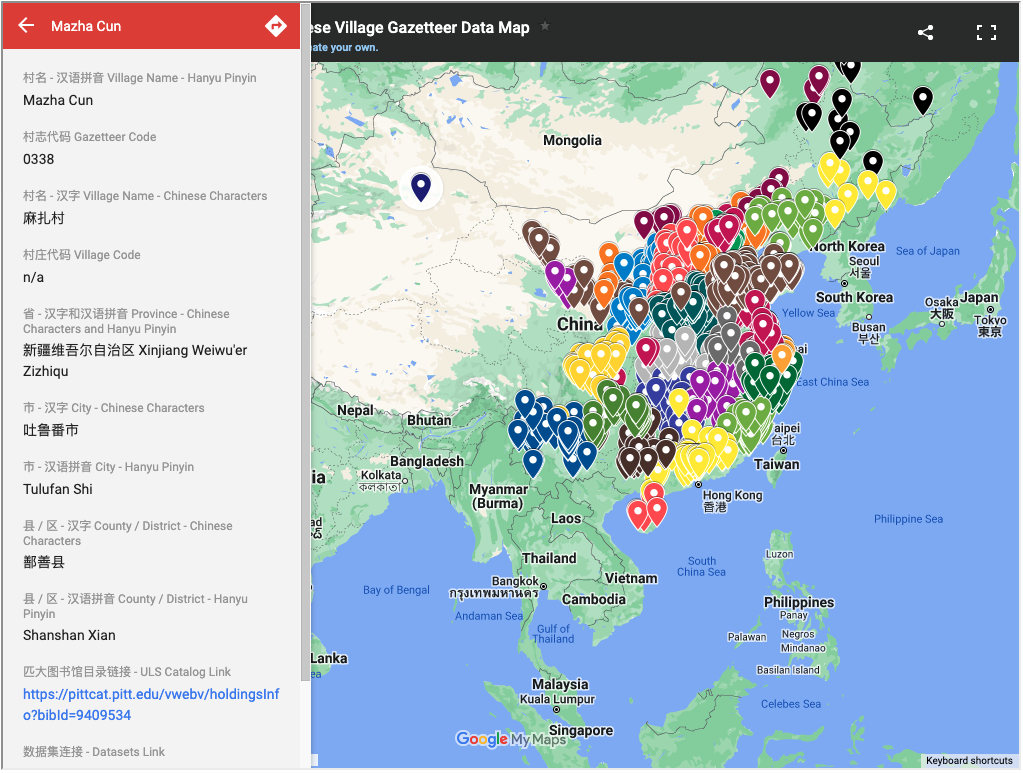}
 \caption{Demonstration of the Map Visualization on CCVG data platform.} 
 \label{fig:2}
\end{figure}

For the CCVGD platform, visualization is one important feature aiming to guide users in their data exploration. Currently, two major forms of visualization are being developed. First is map, which utilizes the contemporary map of China as a base map and marks all the villages covered in the data platform (see \autoref{fig:1}). Built with Google Maps, this interactive map enables users to develop an overview of the geographic distribution of the villages and explore further details of specific villages (e.g., see \autoref{fig:2}). Another important visualization form to highlight data content is graphs and charts (e.g., line graphs, bar charts) reactive to users’ data search on the platform, providing a visual tool to explore and analyze the massive numeric data. This step is still in the design stage, with experimentation using population and economic data for selected villages. \par

\begin{figure}[tb]
 \centering 
 \includegraphics[width=\columnwidth]{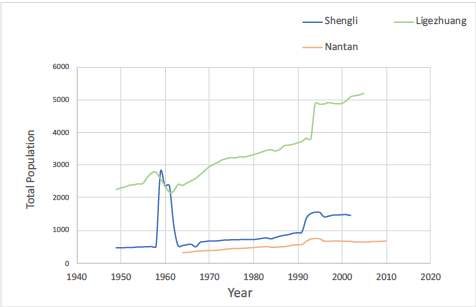}
 \caption{Total Population of Shengli, Ligezhuang, and Nantan Villages from 1949 to 2010.} 
 \label{fig:3}
\end{figure}

\begin{figure}[tb]
 \centering 
 \includegraphics[width=\columnwidth]{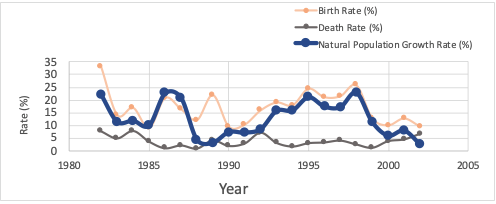}
 \caption{Birth Rate, Death Rate, and Natural Population Growth Rate for Shengli Village between 1980 and 2005.} 
 \label{fig:4}
\end{figure}

The following is an example where I demonstrate visualization design with population data of Shengli, Ligezhuang, and Nantan Villages in Shandong Province. The population data include four major categories: (1) basic population information covering the total population, male population, female population, and the number of households, (2) migration and population change data including migration-in (\textit{qianru}) and migration-out (\textit{qianchu}) data, temporary population, and change of residency status, (3) natural population change data including number of births and deaths, birth rate, death rate, and natural population growth rate, and finally, (4) family planning data covering a variety of sub-categories such as family planning rate, number of women of childbearing age, number of women who have contraception, abortions, and the use of different contraception methods (e.g., tubal ligations, IUD). \autoref{fig:3} and \autoref{fig:4} showcase some of the visualizations under design. Overall, visualizations are designed to support two major search functions of the data platform, the single-village and multi-village search. Single-village search, as the name suggests, is designed to help researchers explore specific data points on a single village. In comparison, multi-village search aims to facilitate comparison across villages and the analysis of broader trends and patterns.\par

Depending on users’ search activities, the predesigned, reactive visualizations are displayed on the platform to highlight data patterns, offering an interactive way to explore characteristics of data, compare data of multiple villages, as well as to observe any irregularities or inconsistency of data. For example, \autoref{fig:3} shows the total population from 1949 to 2010 among Shengli, Ligezhuang, and Nantan Villages. Data differences as well as the sudden drop around 1960s raise potentially interesting questions about the three villages. Visualizations in this project also provide a way to detect accuracy and authenticity issues of data. Due to the unofficial source of data and sometimes individually oriented data collection and compilation process of village gazetteers, data quality issues are intrinsically an important problem of investigation for such materials. Take the Shengli Village as an example. As shown in \autoref{fig:4}, the trend of natural population growth rate between 1987 and 1988 outnumbers the differences between birth and death rates, demonstrating abnormalities and errors in data. Building on the observations of the visualizations, a closer check on the data presents possible collection or calculation errors for the data compiled between 1982 and 1989. Having a visual overview of the datasets before downloading them for further analysis, researchers will be able to develop a critical idea of the datasets at a macro level. \par

The strong abstraction and simplification power that visualizations present enhanced the representation of numeric village data. However, despite the various benefits, current visualization design and methods cannot address every need and demand for potential research. Based on preliminary feedback from humanities researchers and potential users for the data platform, we realized that details and contexts (e.g., texts, narratives, and images in village gazetteers) are equally, if not more important to researchers. Visualizations in this case, however, guide users away from such detailed contextual information by emphasizing on ``big pictures.'' To address this limitation of the current visualization design, more information needs to be extracted from the published gazetteers. More particularly, due to copyright regulations, full-text gazetteers cannot be digitized and included in the data platform, adding infrastructural challenges to implement the most effective visualizations for the project. Shneiderman's theory of ``details on demand" \cite{shneiderman_eyes_1996} offers the technical solution for the visualization dilemma, but the constraints on raw data and materials, which are frequently present in digital humanities projects, also pose unique challenges to the design and application of visuals in digital humanities work. \par

As the purpose of the CCVGD project is to create and provide unique resources for scholars, visual literacy is another important factor to consider when proposing visualization solutions. The increase of information visualizations asks for a higher level of visual literacy and contributes to the challenges of interpreting the visualizations from the users’ side. As the data platform is going to be facing a wide pool of users, how to make visualizations accessible and valuable enough for everyone to use while at the same time informative, accurate, and of high quality is an important problem to evaluate. During the current phase of the project, we do not have a good answer and solution to this question, which can benefit from further discussions.\par

\subsection{Visualizing Video Oral History Data}

Visualization attempts for the CCVGD project highlight the power of graphs and charts in revealing the simplified, macro-level data patterns for digital humanities projects. In this following section, I reflect on another case where visual representations are used in different manners however have also enhanced the data representation. More specifically, I use the Cultural Revolution 10 project to explore the power of visual representations beyond the forms of graphs and charts, particularly the power to organize and shape narratives.\par

The Cultural Revolution 10 (i.e., CR/10) project is an oral history project collecting and preserving video accounts of China’s Cultural Revolution (1966-1976). Initiated by the University of Pittsburgh in 2016, CR/10 (~\href{https://culturalrevolution.pitt.edu/}{https://culturalrevolution.pitt.edu/}) aims to explore the shift of memory accounts on China's Cultural Revolution through hundreds of 10-minute video interviews with participants across generations and backgrounds, who have either experienced the historical incident or learned about it from external resources \cite{ward_collecting_2018,ma_curating_2022}.\par

\begin{figure}[tb]
 \centering 
 \includegraphics[width=\columnwidth]{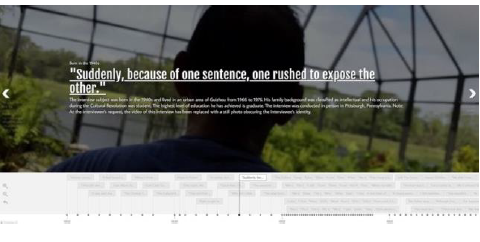}
 \caption{Timeline for CR/10 project.} 
 \label{fig:5}
\end{figure}

\begin{figure}[tb]
 \centering 
 \includegraphics[width=\columnwidth]{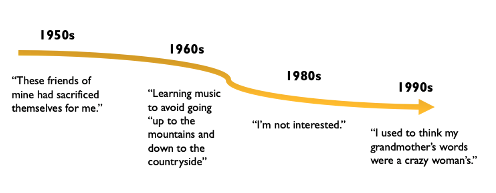}
 \caption{Illustration of the “Atlas of Forgetting” created by the Interactive Timeline.} 
 \label{fig:6}
\end{figure}

\begin{figure}[tb]
 \centering 
 \includegraphics[width=\columnwidth]{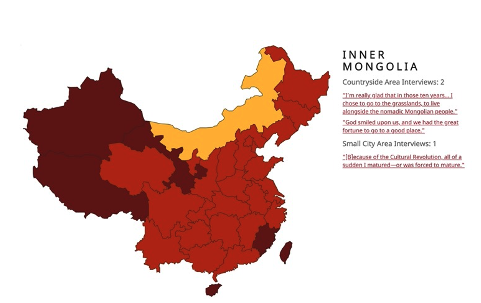}
 \caption{Map Visualization for CR/10 Project.} 
 \label{fig:7}
\end{figure}

Two major forms of interactive visualization are designed and incorporated in the project website, which are used as \textit{navigation paths} for the oral history video collection (see \autoref{fig:5}, \autoref{fig:6}, and \autoref{fig:7}). The timeline visualizes the individual video units based on the interviewee’s date of births (\autoref{fig:5} and \autoref{fig:6}). This form of visualization creates an ``atlas of forgetting" (\autoref{fig:6}), from which viewers of the collection will experience how the cultural memories of China’s Cultural Revolution are gradually fading among generations \cite{ma_curating_2022}. In addition to the timeline, the interactive map (\autoref{fig:7}) offers a way to explore the collected oral history accounts from a geographical perspective. The result generates a powerful narrative of the different degrees of impact the Cultural Revolution has posed to various regions. \par

As described above, none of the visual representations are what we usually refer to as ``visualizations," which essentially focus on quantitative displays of information such as charts and graphs. However, it is undeniable that the embedded visuals, including the interactive timeline, map, and narrative illustrations, have played an invaluable role in transforming the video collection into an argumentative artifact, something Aby Warburg may describe as a cultural atlas \cite{johnson_memory_2012}. User studies research have also demonstrated that the three visual pathways assist with users, especially users who are new to the subject of China's Cultural Revolution, in interpreting individual videos as well as the meaning of the CR/10 project \cite{ma_curating_2022}. To some extent, this case study brings back the notion of \textit{direct visualization} that Manovich discussed \cite{manovich_what_2011}. Humanities datasets of various formats usually contain rich and essential details about the subjects, and therefore, keeping the original forms of data during the process of creating visualizations could add value to the subject and the project. The visualization efforts of this project further push forward the boundaries in thinking about \textit{what is the meaningful visualization for humanities and cultural data}. \par



\subsection{Digitization as a Form of Visualizing Historical Archival Data}

\begin{figure}[tb]
 \centering 
 \includegraphics[width=\columnwidth]{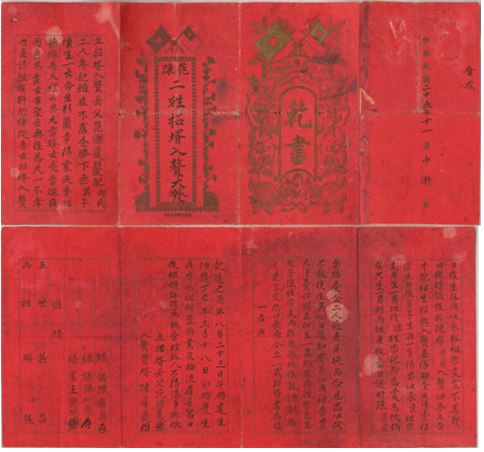}
 \caption{Example of a Digitized Chinese Marriage Record.} 
 \label{fig:8}
\end{figure}

To further extend this discussion, I discuss a final case in this section, which asks: Can high-fidelity digitized images also function as powerful visual representations (or a form of humanities visualizations) for a collection of historical Chinese marriage records? The Chinese marriage records collection spans the period of 1909-1997 and includes a wide range of document types. As an important genre of historical archives, Chinese marriage records are invaluable research sources and the digitization process made it more accessible to the wider public, adding to the social value of the materials \cite{ma_translational_2020}. \par

The following figure is a demonstration of the digitized images for the collection (see \autoref{fig:8}). The high-fidelity digitized images for the collection displayed online served as a medium for the curators to highlight important and unique features of the materials and for the readers to explore and discover them in a more visually accessible and enhanced manner. In this sense, the digitized images function as effective visual representations of the archival data in this collection. Compared with other forms of visual representations discussed in the other two cases, this form of visual representation puts even less emphasis on discovering patterns or structuring materials and focuses more highlighting the rich details and features of original documents.\par

Although compared with visual representations in the other two cases, high-fidelity digitized images have maintained rich and culturally interesting information, they are not perfect, particularly in terms of their ability to retain materiality of the archival collection. Thinking in the context of this collection, historical Chinese marriages have strong social functions where the ways in which they are folded, passed on from one family to another, and circulated and witnessed among the local community as an agreement and contract fulfill and define the meaning of the archives. But these aspects of the historical records cannot be effectively represented in the digitized images; the sole focus on the digitized images of the records would easily mislead the readers to neglect the social functions of the records in the historical time. Given this situation, how can we design visual representations that also engage with the \textit{contexts} and \textit{materiality} of cultural objects? Would augmented or virtual reality be an option? Furthermore, how do curators and cultural heritage institutions balance the visualization quality and infrastructural limitations? Such visualization ideas were not implemented for the Chinese marriage records collection due to practical and infrastructural constraints, but the efforts and reflections in this case study may serve as a preliminary step to examine advanced visualization options in digital archive platforms and infrastructures.\par

\section{CONCLUSION}

In this short paper, I reflect on the roles and functions of visualizations, and more broadly, visual representations, for humanities and cultural data. I have discussed three case studies of distinct nature, namely, the Contemporary Chinese Village Gazetteer Data project, the China’s Cultural Revolution oral history project, and the digitization of historical Chinese Marriage Records project. Despite the different data types, project stages, and project purposes, all the three cases utilize visual methods and technologies as an important method to enhance the representation of humanities and cultural data. By comparing the use of visual representations in these three cases, I reflected on the connotations and boundaries of \textit{visualizations} within the context of humanities data and projects. The multiple modalities and various characteristics of humanities data have raised the potential of expanding the understandings of \textit{visualization} and appreciating different forms of visual representations. Discussion of the three cases also proposed new questions pertinent to humanities visualizations, which deserve further reflections. For example, visualizations implemented in real projects are often the products of compromise between the ideal design and the constraints of available knowledge infrastructures. How do we balance and create functional, ``compromised'' visual representations while appreciating, evaluating, and improving them? In addition, since all the three cases discussed in the paper work with culturally oriented materials and documents, another interesting question to think about is if visualizations are \textit{cultural}, rather than \textit{natural}, and how cultural elements influence the design and application of visual representations. 


\bibliographystyle{abbrv-doi}

\bibliography{template}
\end{document}